\documentclass[
 amsmath,
 amssymb,
 preprint
]{revtex4-1}

\usepackage[utf8]{inputenc}
\usepackage{qcircuit}
\usepackage{physics}
\usepackage{graphicx}
\usepackage{dcolumn}
\usepackage{bm}
\usepackage[export]{adjustbox}
\usepackage{mathtools}
\usepackage{float}
\usepackage{hyperref}
\usepackage{xcolor} 
\usepackage[noabbrev,capitalize]{cleveref}
\usepackage{abraces}
\usepackage{makeidx}
\usepackage{rotating}
\usepackage{verbatim}
\usepackage{mathrsfs}
\usepackage[utf8]{inputenc}
\usepackage{amsmath,amssymb}
\usepackage{hyperref}
\usepackage{algorithm}
\usepackage{algpseudocode}
\usepackage{subcaption}

\DeclarePairedDelimiter{\ceil}{\lceil}{\rceil}

\newcommand\citess[1]{\citeauthor{#1}'s \cite{#1}}

\begin{document}

\title{A quantum walk assisted approximate algorithm for bounded NP optimisation problems}

\author{S. Marsh}
 \email{samuel.marsh@research.uwa.edu.au}
 \affiliation{School of Physics, University of Western Australia}
\author{J. B. Wang}
 \email{jingbo.wang@uwa.edu.au}
 \affiliation{School of Physics, University of Western Australia}

\date{\today}

\begin{abstract}
This paper describes an application of the Quantum Approximate Optimisation Algorithm (QAOA) to efficiently find approximate solutions for computational problems contained in the polynomially bounded NP optimisation complexity class (NPO PB). We consider a generalisation of the QAOA state evolution to alternating quantum walks and solution-quality-dependent phase shifts, and use the quantum walks to integrate the problem constraints of NPO problems. We apply the recent concept of a hybrid quantum-classical variational scheme to attempt finding the highest expectation value, which contains a high-quality solution. The algorithm is applied to the problem of minimum vertex cover, showing promising results using only a fixed and low number of optimisation parameters.
\end{abstract}


\maketitle

\section{Introduction}

Quantum computers exploit the properties of quantum mechanics such as superposition and entanglement, providing the ability to solve certain computational problems~\cite{Shor:1997bl,Childs2002,Grover1996} far more efficiently than any classical computer. However, the power of quantum computation does not apply indiscriminately to all computational problems. It is an active area of study as to whether a quantum advantage applies to the class of `NP optimisation problems'. An \textit{NP optimisation problem}~\cite{Kann:1995kx} can be defined by a four-tuple $(I, s, c, g)$, such that
\begin{itemize}
\item $I$ is the set of problem \textit{instances} (specific cases of the general abstract problem). It must be efficient to determine if a particular object belongs to $I$.
\item For each instance $x \in I$, the function $s : I \rightarrow \mathcal{P}(\mathcal{U})$ (with $\mathcal{U}$ representing the universal set) returns the set of valid, or feasible, solutions to $x$. The size $|y|$ of any solution $y \in s(x)$ must be bounded from above by some polynomial function of the size $|x|$ of $x$. In addition, given $x \in I$ and $y$ with $|y|$ bounded from above by some polynomial in $|x|$, it must be efficiently verifiable as to whether $y \in s(x)$.
\item For $x \in I$ and $y$, $c : I \times \mathcal{U} \rightarrow \mathbb{Z}^*$ is the objective function, or \textit{measure}. This function returns a non-negative integer representing the quality of the solution $y$ with respect to $x$, and is efficiently computable. The function $c$ only returns a meaningful result when $y \in s(x)$.
\item $g$ is the goal function, either \textit{max} or \textit{min}.
\end{itemize}
Given problem instance $x$, the aim of a NP optimisation problem is to find $y$ such that
\begin{equation}
    c(x, y) = g(\{ c(x, y') : y' \in s(x) \}).
\end{equation}
The complexity class NPO labels the set of all NP optimisation problems.

The polynomially-bounded NP optimisation problem class (NPO PB) adds the further restriction of $c$ being bounded by some polynomial function in the size of the problem instance~\cite{Kann:1995kx}. Many optimisation problems such as minimum vertex cover, graph partitioning, and maximal clique are contained in NPO PB~\cite{Lucas:2013eb}. Optimisation problems such as integer programming, number partitioning and travelling salesman are not contained in NPO PB~\cite{Kann1993}. For real-world application of NPO problems, an approximate algorithm can suffice, which aims to find a good solution efficiently.

In 2014, \citet{Farhi2014} published a new algorithmic framework called the Quantum Approximate Optimisation Algorithm (QAOA) for finding approximate solutions to combinatorial optimisation problems using quantum computation. This framework was applied to some example optimisation problems, returning `good' solutions according to the relevant metric. For a combinatorial optimisation problem with integer objective function $c$ where solutions can be encoded using $n$ bits, they define a diagonal quantum operator $\hat{C}$ by its action on the $n$-dimensional computational basis states $\ket{x}$ such that $\hat{C} \ket{x} = c(x) \ket{x}$. The authors also define an operator $\hat{\mathcal{B}} = \sum\limits_{i = 1}^n \sigma^x_i$,
where $\sigma^x_i$ is the Pauli-X operator acting on the $i$th qubit of the register. The authors then make use of the quantum adiabatic theorem~\cite{Farhi2000}. Since $\hat{\mathcal{B}}$ satisfies the Perron-Frobenius requirements, by evolving a system initially in the highest-eigenvalue eigenstate of $\hat{\mathcal{B}}$ under the influence of a Hamiltonian which slowly interpolates from $\hat{\mathcal{B}}$ to $\hat{C}$ over a large time $T$, the final state of the system will be the highest-eigenvalue eigenstate of $\hat{C}$. Taking the linear interpolation
\begin{align}
    & \hat{H}(t) = \frac{t}{T} \hat{C} + (1 - \frac{t}{T}) \hat{\mathcal{B}}, & t \in [0, T] &
\end{align}
and performing Trotterisation on the time evolution into $p$ timesteps followed by a further Trotterisation on each of the resultant terms leads to the state evolution
\begin{equation}
    \ket{\vec{\beta}, \vec{\gamma}} = e^{-i \beta_p \hat{\mathcal{B}}} e^{-i \gamma_p \hat{C}} \ldots e^{-i \beta_1 \hat{\mathcal{B}}} e^{-i \gamma_1 \hat{C}} \ket{s} .
\end{equation}
The state $\ket{s}$ is the $n$-dimensional equal superposition, corresponding to the highest-eigenvalue eigenstate of this particular $\hat{\mathcal{B}}$. The $2p$ unknowns $\vec{\beta} = (\beta_1, \ldots \beta_p)$ and $\vec{\gamma} = (\gamma_1, \ldots, \gamma_p)$ are treated as optimisation parameters, with the optimal values corresponding to an evolution path that replicates that of $\hat{H}(t)$ as closely as the parameter space allows. The search space can be restricted to $\gamma \in [0, 2\pi)^p$ and $\beta \in [0, \pi)^p$ because both $\hat{C}$ and $\hat{\mathcal{B}}$ have integer eigenvalues. The QAOA takes the optimal parameter values which maximise the expectation value $F_p(\vec{\beta}, \vec{\gamma}) = \bra{\vec{\beta}, \vec{\gamma}} \hat{C} \ket{\vec{\beta}, \vec{\gamma}}$, since a high expectation value with respect to $\hat{C}$ means a solution $x$ with a high value of $c(x)$ on average. The QAOA has the critical property that
\begin{equation}
    \lim_{p \rightarrow \infty} \max\limits_{\vec{\beta}, \vec{\gamma}} F_p(\vec{\beta}, \vec{\gamma}) = \max\limits_{x} c(x)
\end{equation}
and
\begin{equation}
    \max_{\vec{\beta}, \vec{\gamma}} F_p(\vec{\beta}, \vec{\gamma}) \geq \max_{\vec{\beta}, \vec{\gamma}} F_{p-1}(\vec{\beta}, \vec{\gamma}).
\end{equation}
Consequently the algorithm's performance improves with $p$, guaranteeing the optimal solution in the limit. \citeauthor{Farhi2014} then restrict to very low $p$, choosing to study $p = 1$ for the NP optimisation problem of maximum cut.

In this paper, we consider a generalisation of the QAOA state evolution as a series of quantum walks interleaved with solution-quality-dependent phase shifts, and use the quantum walks to integrate the problem constraints of NPO problems. A continuous time random walk on a graph $G$ models the flow of probability between neighbouring vertices on the graph. This concept was extended to the quantum domain by \citet{Farhi1998}. Consider a graph $G = (V, E)$ with adjacency matrix $A$. For our purposes it is convenient to assume that $G$ has $2^n$ vertices, so the vertices can be identified with the $2^n$-dimensional computational basis states. Then the continuous time quantum walk on $G$ can be defined by the propagator $\hat{U}(t) = e^{-i t \hat{A}}$ with respect to a $n$-qubit quantum register, where $\hat{A}$ is the $2^n$-dimensional quantum operator defined on the computational basis states by the adjacency matrix $A$. The probability distribution over the graph after time $t$ is held in the probability of measuring each of the basis states after the action of operator $\hat{U}(t)$ on the initial quantum state. In contrast to the classical random walk, interference and other quantum phenomena can come into effect. This leads to markedly different behaviour between the two cases. Continuous time quantum walks provide a versatile platform for universal quantum computation~\cite{Childs791}. They have been used extensively in graph theoretical applications~\cite{Izaac2017,Loke2017,Gamble2010,Chimera} and are the basis of many other quantum algorithms~\cite{Spatialsearch,ZhiJian2015,QWBook,Qiang2016}.

The QAOA also requires an efficient method for determining a specific problem-dependent expectation value. We adopt the recent concept of a hybrid quantum-classical variational scheme~\cite{Peruzzo2014} for this purpose, and prove its efficiency for all problems in NPO PB. These results are aggregated to present a final quantum algorithm for finding approximate solutions to any problem in the NPO PB class, using the minimum vertex cover problem as a representative example. A vertex cover of a graph $G = (V, E)$ is a subset of the vertices such that for every $\langle u  v \rangle \in E$, either $u$ or $v$ is in the set. That is, every edge has at least one end in the set. Out of all vertex covers existing for $G$, the minimum vertex cover is the one with the fewest vertices. This is an NPO PB problem~\cite{Lucas:2013eb}, with the goal to maximise the number of vertices \textit{not} in the vertex cover. As well as being a useful problem to state and study in terms of computational complexity, minimum vertex cover has wide applicability to real world problems~\cite{worms,Lancia:2001im,OCallahan:2003bt,Hou:1994jp,Domingo1999}. The algorithm is shown to produce high-quality solutions efficiently for various classes of minimum vertex cover problem instances.

In the following section, we demonstrate how the constraints associated with NPO problems can be encoded into the QAOA framework. Next, we give an efficient strategy for finding the optimal QAOA parameters for NPO problems with polynomially bounded measure. We show that the overall algorithm has an efficient quantum circuit. Finally, we give results for small instances of minimum vertex cover.
 
\section{Encoding NPO problems}

Consider the maximum-size solution $x \in s(y)$ for problem instance $y \in I$ of an NPO problem $(I, s, c, g)$. Since $|x|$ is polynomially bounded in $|y|$, it must be possible to encode $x$ in some unique binary string of length $n$, with $n$ growing at most polynomially in $|y|$. This binary string can be represented as a decimal number ranging from $0$ to $2^n - 1$. Given that $x$ is the maximum-size solution by definition, all other solutions $x' \in s(y)$ can also be represented by unique length-$n$ binary strings.

Consequently, it suffices to consider the integers in the range $0 \ldots 2^n - 1$. Some of these integers will correspond to feasible solutions, while others may not. $s : I \rightarrow \mathcal{P}(\mathcal{U})$ can be redefined as $s : I \rightarrow \mathcal{P}(\mathbb{Z}^*)$, returning for an instance $y \in I$ a set of integers in the range $0 \ldots 2^n - 1$, for some integer $n$ and corresponding to some set of unique feasible solutions. It can also be assumed without loss of generality that $g = \textit{max}$. In addition the shorthand $c(x, y) \equiv c(x)$ will be used, with the implicit understanding that $c$ may depend on the specific problem instance.

Given this, there is a natural way to encode the measure of any NPO problem $(I, s, c, g=\textit{max})$ into the QAOA, simply defining $\hat{C}$ by
\begin{equation}
    \label{eq:npoc}
    \hat{C} \ket{x} = c(x) \ket{x},
\end{equation}
with $c$ the measure of the NPO problem. Without modification, the QAOA will work to produce solutions $x$ with a high value of $c(x)$. However, these produced solutions need to be feasible, such that $x \in s(y)$ for a given problem instance $y \in I$.

\begin{figure}[b]
\centering
\includegraphics[width=0.45\linewidth,valign=m]{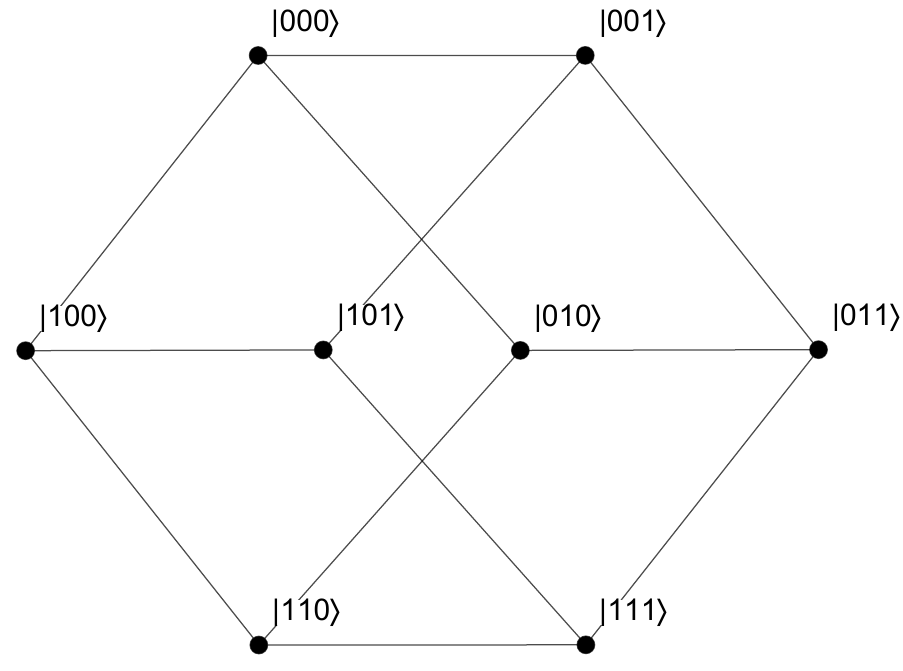}
\includegraphics[width=0.45\linewidth,valign=m]{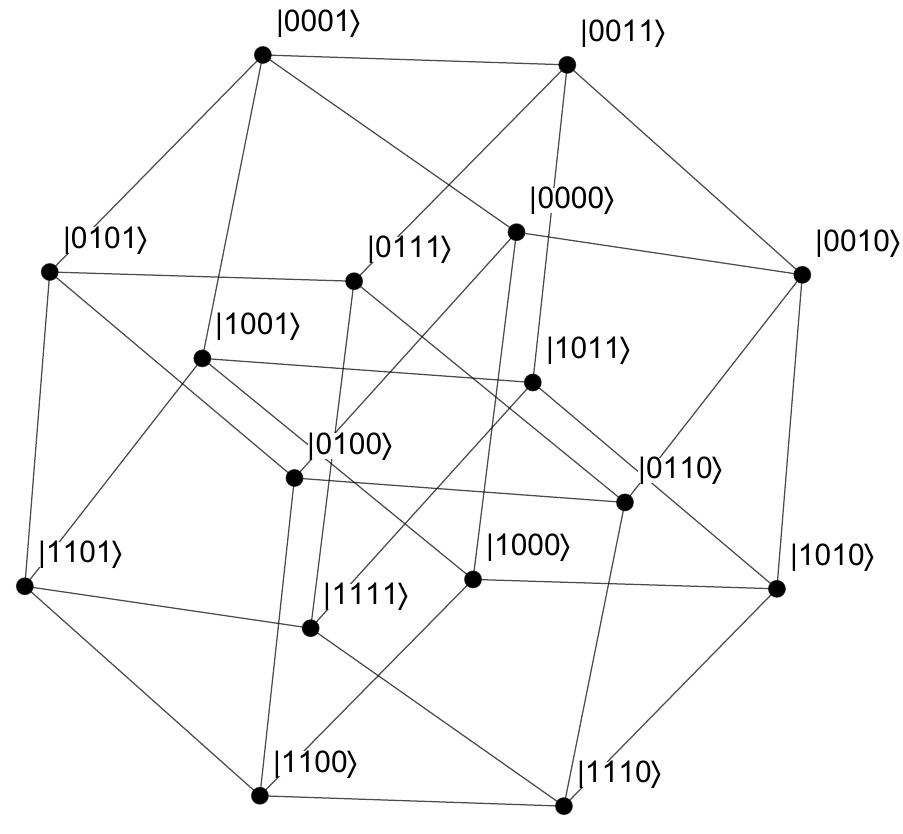}
\caption{Representation of the transverse field operator $\hat{\mathcal{B}}$ as an $n$-dimensional hypercube in three and four dimensions respectively.}
\label{fig:unconstrainedb}
\end{figure}

In order to enforce this, the structure of $\hat{\mathcal{B}}$ needs to be considered. The transverse field operator $\hat{\mathcal{B}}$ can also be defined equivalently in terms of matrix elements:
\begin{equation}
    \bra{x} \hat{\mathcal{B}} \ket{x'} = \begin{cases}
        1   &   \text{$x$ and $x'$ differ in a single bit,} \\
        0   &   \text{otherwise.}
    \end{cases}
\end{equation}
This definition reveals additional structure of $\hat{\mathcal{B}}$: it represents the adjacency matrix of a hypercube. Each of the vertices of the hypercube are associated with a computational basis state $\ket{x}$ with edges connecting basis states $\ket{x}$ and $\ket{x'}$ if $\bra{x} \hat{\mathcal{B}} \ket{x'} = 1$, as per \cref{fig:unconstrainedb}.

\begin{figure}[H]
\centering
\begin{subfigure}[t]{0.45\linewidth}
\centering
\includegraphics[width=\linewidth,valign=m]{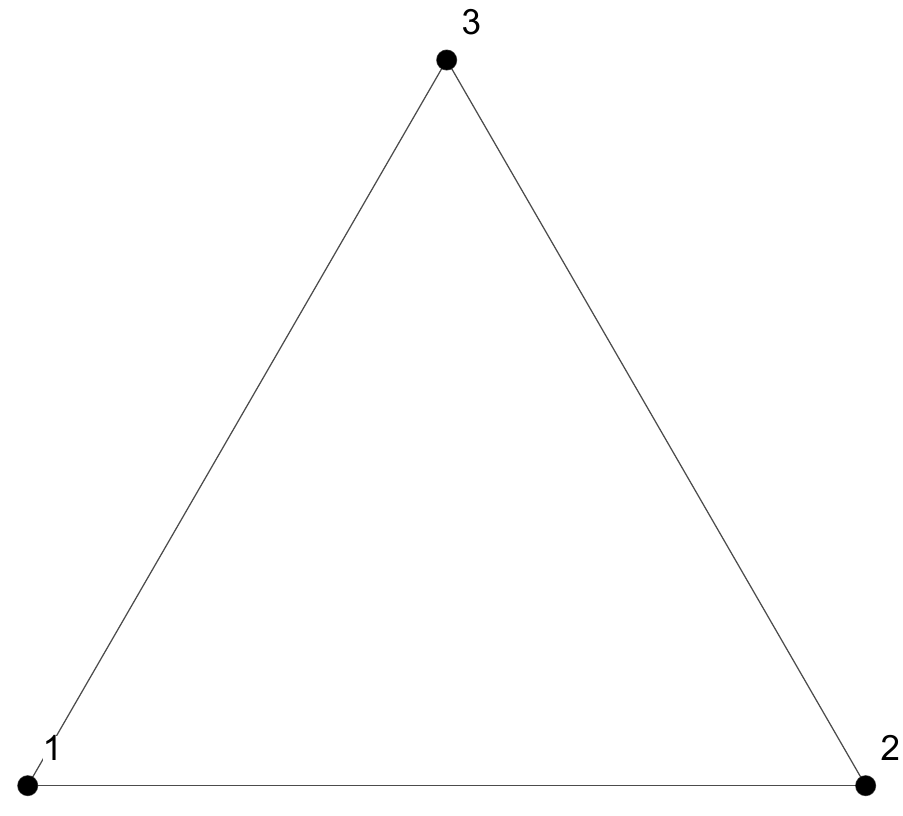}
\caption{}
\label{fig:3vertexgraphexample}
\end{subfigure}
\begin{subfigure}[t]{0.45\linewidth}
\centering
\includegraphics[width=\linewidth,valign=m]{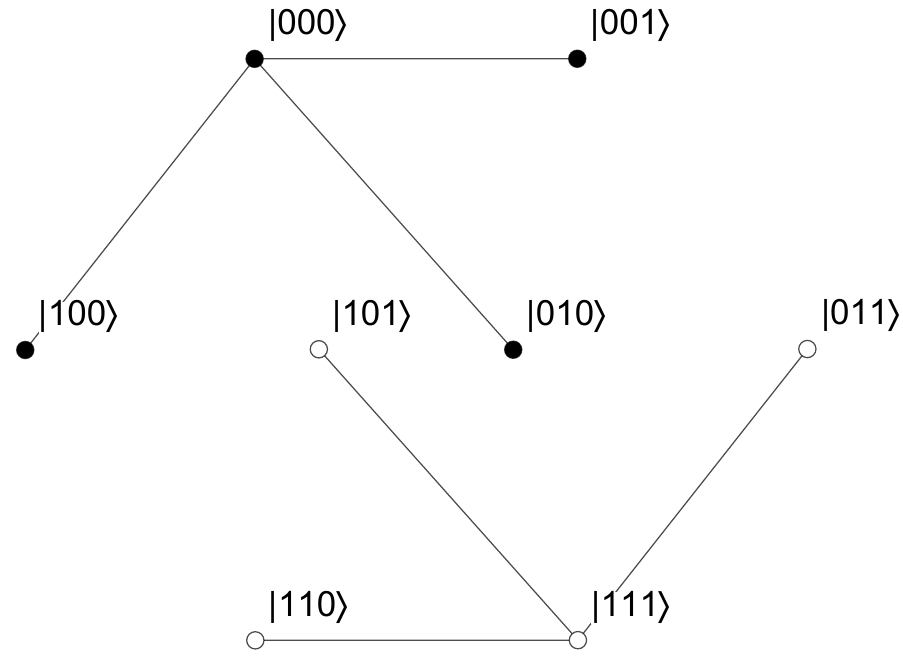}
\caption{}
\label{fig:constrainedbexample}
\end{subfigure}
\caption{(a) shows an arbitrary three-vertex input graph, for which the aim is to find the minimum vertex cover. (b) depicts the corresponding $\hat{B}$ operator for this graph, with states representing vertex covers (open circles) and states not representing vertex covers (filled circles) separated into two disjoint subgraphs.}
\label{fig:exampleconstrainedb}
\end{figure}

For any NPO problem $(I, s, c, g)$ it is efficient by definition to compute if $x \in s(y)$ given problem instance $y$. Hence a `validation function' $v$ can be defined, such that
\begin{equation}
    \label{eq:validation}
    v(x) = \begin{cases}
        1 & \text{$x$ is a feasible solution,} \\
        0 & \text{otherwise.}
    \end{cases}
\end{equation}
The function value $v(x)$ is efficiently computable for all $x \in 0, \ldots, 2^n - 1$. Then defining a modified $\hat{\mathcal{B}}$ operator, $\hat{B}$, we can incorporate problem constraints as follows:
\begin{equation}
    \label{eq:constrainedb}
    \bra{x} \hat{B} \ket{x'} = \begin{cases}
        1   &   \text{$x$ and $x'$ differ in a single bit and $v(x)=v(x')$,} \\
        0   &   \text{otherwise.}
    \end{cases}
\end{equation}
This acts as a disconnection of the hypercube into two disjoint subgraphs -- one containing the feasible solutions, and the other containing the infeasible solutions which do not satisfy problem constraints. $\hat{B}$ still satisfies the Perron-Frobenius requirements and thus $\hat{H}(t)$ will continue to satisfy the adiabatic conditions~\cite{Farhi2000}, so the theory behind the QAOA is still valid in this regard. \cref{fig:exampleconstrainedb} presents a vertex cover-specific example of the $\hat{B}$ operator.

The unitary operator $e^{-i \beta \hat{B}}$ represents a continuous time quantum walk~\cite{Farhi1998} over the feasible states of the adjacency matrix $\hat{B}$. This adjacency matrix represents a modified hypercube, where vertices correspond to unique solutions to the combinatorial optimisation problem in question. By modifying the hypercube operator $\hat{B}$ to disconnect the feasible solutions from the infeasible solutions, and modifying the initial state $\ket{s}$ to have non-zero probability only for states representing feasible solutions, the  quantum walk via $e^{-i \beta \hat{B}}$ will never `enter' an infeasible state. In combination with the fact that $e^{-i \gamma \hat{C}}$ is a diagonal unitary and so does not modify state amplitudes, this means that the state $\ket{\vec{\beta}, \vec{\gamma}}$ (with an appropriately modified initial state) will always guarantee a feasible solution when measured.

It is required that the highest-energy state of $\hat{B}$ be known and efficiently-preparable, in order to perform an adiabatic evolution starting from this state~\cite{Farhi2000}. The original transverse field (or hypercube) operator $\hat{\mathcal{B}}$ satisfied this requirement, with the highest-energy state being the equal superposition state. However, incorporation of constraints via $\hat{B}$ means that the highest-eigenvalue state is no longer the equal superposition. Rather, it will be some non-trivial superposition over the states dependent on the structure of the constraints and the problem instance -- not at all easy to find, let alone prepare efficiently. To circumvent this issue, a `prior' adiabatic evolution can be performed. This is done in~\cite{Farhi2014} for the specific NPO problem \textit{maximum independent set}. The first evolution is from the highest-eigenvalue state of $-\hat{C}$ to the highest-eigenvalue state of $\hat{B}$. The highest-eigenvalue state of $-\hat{C}$ is equivalent to the ground state of $\hat{C}$, corresponding to the lowest-quality solution to the NPO problem. For many NPO problems, this lowest-quality solution can be found efficiently. For example, the lowest-quality vertex cover of a graph corresponds to the cover using every one of the $n$ available vertices. However, if the lowest-quality solution cannot be found efficiently, the algorithm supports a generalisation to the use of \textit{any} feasible solution as the initial state. This inspires a  transition from an adiabatic perspective (requiring the initial state to be the lowest-quality feasible solution) to a quantum walk perspective (supporting any feasible state as the initial state). The modified QAOA state evolution is described below.

Given a particular NP optimisation problem, assume that an initial feasible solution $\ket{s}$ can be efficiently found and prepared. Then for level-$p$ QAOA, define $2p-1$ parameters $\vec{\beta} = (\beta_1, \ldots, \beta_p) \in \mathbb{R}^p$ and $\vec{\gamma} = (\gamma_1, \ldots, \gamma_{p - 1}) \in [0, 2\pi)^{p - 1}$. With these parameters, the state evolution is defined as the alternating series of operators
\begin{equation}
    \label{eq:newevolution}
    \ket{\vec{\beta},\vec{\gamma}} = e^{-i \beta_p \hat{B}} e^{-i \gamma_{p-1} \hat{C}} \ldots e^{-i \gamma_1 \hat{C}} e^{-i \beta_1 \hat{B}} \ket{s},
\end{equation}
The $e^{-i \beta \hat{B}}$ operator encodes the problem constraints through the modified hypercube operator $\hat{B}$, and performs a continuous time quantum walk over the valid states dependent on the parameter $\beta$. The $e^{-i \gamma \hat{C}}$ operator encodes the NPO measure through the diagonal operator $\hat{C}$, modifying the relative phase of the computational basis states $\ket{x}$ depending on the quality of the solution $c(x)$. The walk operators are the components of the evolution that performs the amplification. Since the high-quality states have a uniquely-distinguished phase due to applications of $e^{-i \gamma \hat{C}}$, the amplitude of these states will be amplified relative to the lower-quality solutions at some point during the quantum walk. Note that the $\vec{\beta}$ parameters can no longer be restricted to $[0, \pi)$ because $\hat{B}$ does not necessarily have integer eigenvalues. The $\vec{\gamma}$ parameters can still be restricted to lie in the range $[0, 2\pi)$, since for any NPO problem the measure $c$ is an integer function, and thus $\hat{C}$ will have integer eigenvalues.

There is one subtlety involved in this method for integrating problem constraints, which forces an additional limit on the computational problems that fit into the algorithmic framework. The assumption made is that by disconnecting the valid solutions from the invalid ones, the subgraph containing the valid solutions is connected. In fact, the restriction is slightly weaker -- there must be a path from the initial state of the algorithm to the solution state (or at least to sufficiently high-quality solutions), for any problem instance. A wide range of NPO problems fit this description, as discussed below.

Take for example the NP optimisation problem of minimum vertex cover. Consider any arbitrary vertex cover represented by bit-string $x$ on a graph $G = (V, E)$. By adding another unused vertex to the cover $x$, the resulting set represented by $x'$ is still a vertex cover: all edges $\langle i j \rangle \in E$ are still covered. In addition, there is an edge in $\hat{B}$ connecting $x$ and $x'$, since they differ in a single bit and both represent vertex covers. This same logic can be applied to $x'$, creating a path of edges from $x$ up to $11 \ldots 1$ (the vertex cover using every vertex). Hence, there is a path along the modified hypercube from every vertex cover to the solution $11 \ldots 1$, and by extension every other vertex cover. So the subgraph of the hypercube operator representing valid vertex covers is connected.

This type of connectivity is a general property of a large number of NP optimisation problems, including set packing, maximum cut, maximum independent set, maximum clique, and hitting set. A typical NP optimisation problem aims to minimise/maximise the number of elements in the set, under some constraint. By adding/removing elements to/from the set respectively, the solution is worsened but still satisfies the constraint. Problems with this property will satisfy the connectivity requirement. The reader is invited to confirm that the examples given, from \citess{Karp2010} original list of 21 NP-complete problems, are some of the problems that fall into this category. It may also be possible to choose $\hat{B}$ differently such that the subgraph remains connected for any problem instance. Recent research has explored the use of various choices of $\hat{B}$ in the QAOA~\cite{Hadfield2017}. 

\section{Finding the optimal parameters}

\begin{figure}
\centering
\includegraphics[width=0.9\linewidth,valign=m]{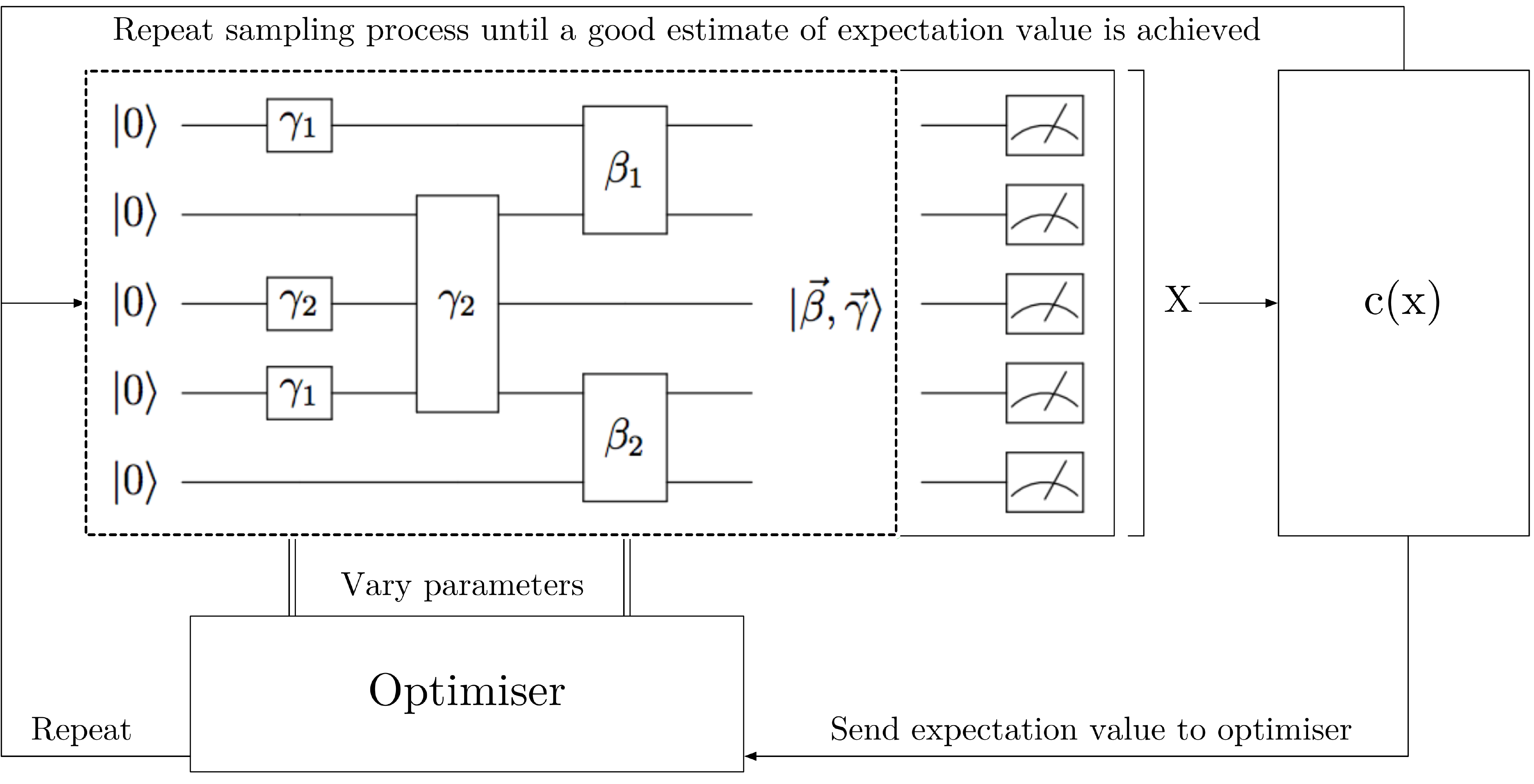}
\caption{Illustration of the hybrid quantum-classical variational method for finding the optimal QAOA parameters $\vec{\beta}^*$ and $\vec{\gamma}^*$. The dashed region is the quantum component.}
\label{fig:hybrid}
\end{figure}

To estimate the expectation value $F_p(\vec{\beta}, \vec{\gamma})$, the state $\ket{\vec{\beta}, \vec{\gamma}}$ is prepared and sampled using the illustrative quantum circuit depicted in \cref{fig:hybrid}. We repeatedly set up the state $\ket{\vec{\beta}, \vec{\gamma}}$, measure the state to obtain a solution bit-string $x$, and then evaluate $c(x)$. The average of these $c(x)$ values  will converge to the expectation value $F_p(\vec{\beta}, \vec{\gamma})$. This estimate can be fed back to the optimiser, and the parameters $\vec{\beta}$, $\vec{\gamma}$ in the quantum state evolution can then be adjusted accordingly. This is the so-called \textit{hybrid quantum-classical} approach as adapted by \cite{Peruzzo2014}.

$F_p(\vec{\beta}, \vec{\gamma})$ can be efficiently found using this method for any NPO PB problem. An NPO PB problem of size $n$ has measure $c$ which is bounded by $[0, c_\text{max}(n)]$, such that $c_\text{max}(n)$ grows at most polynomially in $n$. The expectation value $F_p(\vec{\beta}, \vec{\gamma}) = \bra{\vec{\beta}, \vec{\gamma}} \hat{C} \ket{\vec{\beta}, \vec{\gamma}}$ also lies in this range. According to the central limit theorem, the number of samples required to estimate the mean of a population with variance $\sigma^2$ to within $\epsilon$ is $z^2 \sigma^2/\epsilon^2$, where $z$ is the z-score associated with the required confidence interval~\cite{feller1945}. Using Popoviciu's inequality on variances~\cite{Sharma2010}, for a bounded distribution in $[0, c_\text{max}(n)]$ the variance is at most $\frac{1}{4} c_\text{max}(n)^2$. Hence for a fixed confidence interval the number of samples required grows like $O\left(c_\text{max}(n)^2/\epsilon^2\right)$. This is polynomial with respect to $n$ and is thus an efficient method of finding the expectation value $F_p(\vec{\beta}, \vec{\gamma})$ for any NPO PB problem.

Hence any problem which fits into the NPO PB class is a suitable candidate for this algorithm. With this in mind, the overall variational QAOA process is as follows. Start with some arbitrary initial $\vec{\beta}$ and $\vec{\gamma}$. Repeatedly construct and measure the state $\ket{\vec{\beta}, \vec{\gamma}}$ to get a bit-string $x$, and evaluate $c(x)$. Enough repetitions will give a satisfactory estimate of $F_p(\vec{\beta}, \vec{\gamma})$. Return this value to the optimiser, and obtain a new updated set of parameters $\vec{\beta}$ and $\vec{\gamma}$. This process repeats until a maximum is found, and the optimiser terminates. Throughout this process, keep track of the highest seen value of $c(x)$ and the corresponding bit-string $x$. At the end of the algorithm, this $x$ is taken as the solution.

\section{Efficient quantum circuit}

We now show that this algorithm can be implemented efficiently. There always exists an efficient quantum circuit for the implementation of $e^{-i \gamma \hat{C}}$. \citet{Welch2014} provide a strategy for generation of a quantum circuit to implement this operator without use of ancilla qubits for diagonal $\hat{C}$ having efficiently-computable elements. This builds from work by \citet{Childs:2004}, who proved that if $\hat{C}$ is diagonal and has efficiently-computable elements then an efficient quantum circuit for $e^{-i \gamma \hat{C}}$ can be found. For any NP optimisation problem $(I, s, c, g)$ it is efficient to compute the measure value $c(x)$ for any input $x$ by definition, and thus an efficient quantum circuit can be found.

An efficient quantum circuit for $e^{-i \beta \hat{B}}$ also exists. In \citeyear{Aharonov2003}, \citet{Aharonov2003} proposed a method for efficient implementation of $e^{-i \beta \hat{B}}$ as long as $\hat{B}$ is efficiently row-computable. A Hamiltonian $\hat{H}$ is efficiently row-computable if for every computational basis state $\ket{b}$, all the non-zero matrix elements $\bra{a} \hat{H} \ket{b}$ can be efficiently found.

We can verify that $\hat{B}$ is efficiently row-computable for any NPO problem. Given basis state $\ket{b}$, set $x \gets v(b)$. Then for $i \gets 1, 2, \ldots, n$ toggle bit $i$ of $b$ to produce $a_i$. Set $y_i \gets v(a_i)$. If $y_i = x$, then $\bra{a_i} \hat{B} \ket{b} = 1$. This produces all the non-zero elements of row $a$ of $\hat{B}$ as per \cref{eq:constrainedb}, and makes only $(n+1)$ calls to $v(x)$ which is known to run in polynomial time. Hence $\hat{B}$ is efficiently row-computable. Consequently, $e^{-i \beta \hat{B}}$ always has an efficient quantum circuit. The structure of this circuit is problem-dependent since the constraints affect the non-zero matrix entries of $\hat{B}$. The reader is invited to refer to~\cite{Aharonov2003}, where a method is given for translating an efficiently row-computable operator $\hat{H}$ to the corresponding quantum circuit for $e^{-i t \hat{H}}$. 

In practice, it may be more efficient to instead implement
\begin{equation}
    \bra{x} \hat{B} \ket{x'} = \begin{cases}
        1   &   \text{$x$ and $x'$ differ in a single bit and both $v(x)$, $v(x') \neq 0$,} \\
        0   &   \text{otherwise.}
    \end{cases}
\end{equation}
This has identical behaviour to \cref{eq:constrainedb} in terms of a quantum walk over feasible states, but additionally removes the edges in the `infeasible region' rather than only disconnecting the two regions. This may use fewer gates than \cref{eq:constrainedb} since the matrix $\hat{B}$ has higher sparsity.

\section{Results}

Classical simulations of the quantum state evolution were performed to verify the correctness of the algorithm and to evaluate the quality of approximate solutions in the context of minimum vertex cover. We define the \textit{approximation quality} for a particular problem instance as the ratio of the number of vertices in the minimum vertex cover to the approximate cover. Since for a $n$-qubit quantum register the classical computer must store all $2^n$ quantum amplitudes in memory, results were obtained for only low-$n$ simulations ($\lesssim 20$). A Nelder-Mead non-linear optimiser~\cite{Nelder1965} was used to maximise the expectation value $F_p(\vec{\beta}, \vec{\gamma})$.

An example output of the $p=2$ algorithm is shown in \cref{fig:4veresults}. The maximised expectation value $F_p(\vec{\beta}, \vec{\gamma}) \approx 2.5$ is sufficient for the algorithm to find the optimal solution. This is because in order to have determined the expectation value $F_p(\vec{\beta}, \vec{\gamma})$, the algorithm must have measured at least one solution $x$ with $c(x) \geq F_p(\vec{\beta}, \vec{\gamma})$. The only solutions with this property are the four minimum vertex covers.

\begin{figure}[H]
\centering
\includegraphics[width=0.5\linewidth,valign=t]{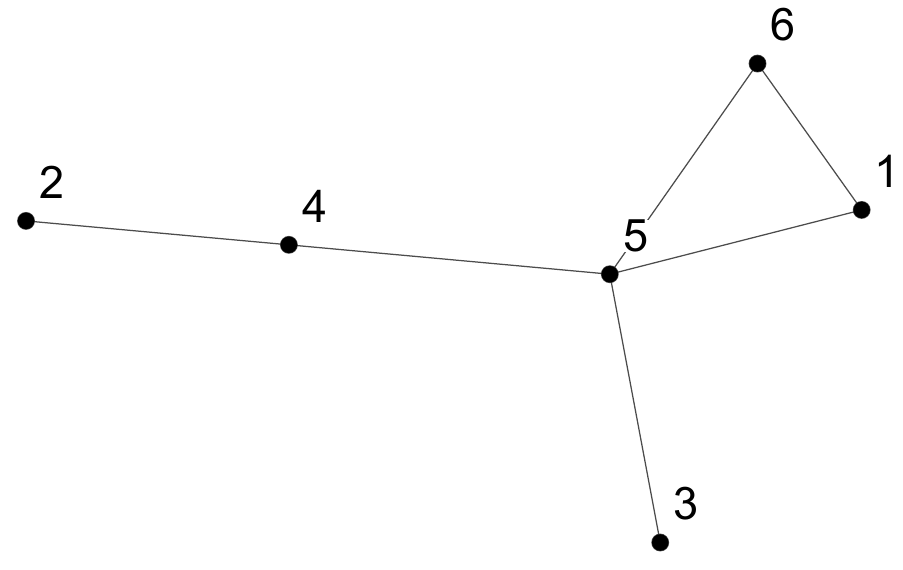}
\caption{An example input graph.}
\label{fig:example_vc}
\end{figure}

\begin{figure}[H]
\centering
\includegraphics[width=\linewidth,valign=t]{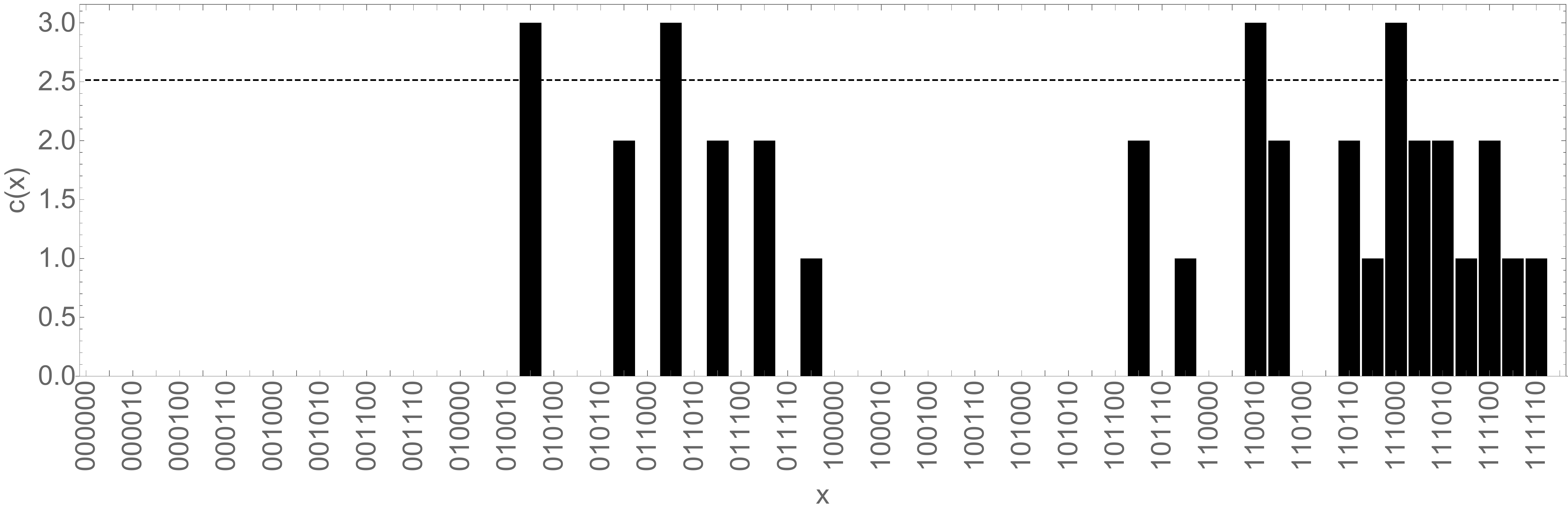}
\caption{Output for the $p=2$ algorithm on the problem instance in \cref{fig:example_vc}, showing the measure for feasible solutions and the maximised expectation value (dashed line).}
\label{fig:4veresults}
\end{figure}

\begin{figure}[H]
\centering
\includegraphics[width=\linewidth,valign=m]{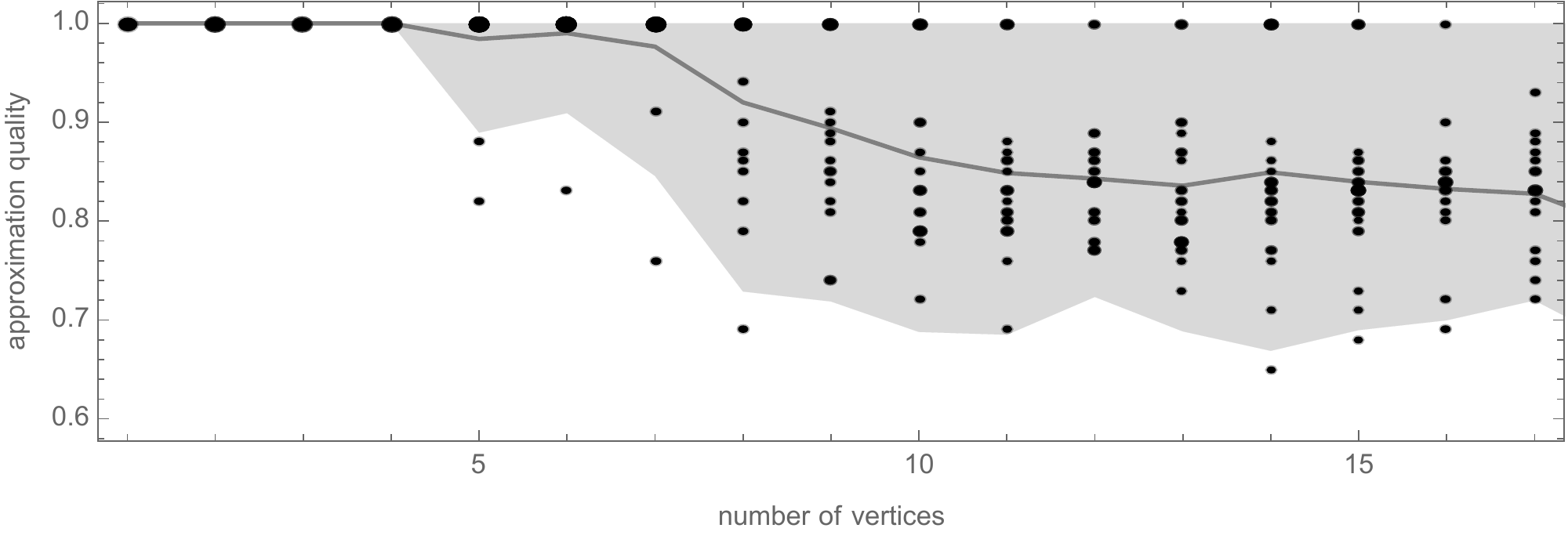}
\caption{Ratio of the number of vertices in the approximate cover to the minimum vertex cover. The grey line is the mean and the shaded region is the 95\% confidence interval. The size of each data point is proportional to the number of tested instances with the same approximation quality.}
\label{fig:erdos}
\end{figure}

The performance of the $p=2$ algorithm was also tested on a random sample of $G(n, 0.5)$ Erdős-Rényi graphs. The $G(n, 0.5)$ Erdős-Rényi random graph model~\cite{Erdos1959} has equal probability to select each of the $2^{n(n-1)/2}$ $n$-vertex graphs, so gives a good impression of the `average case' performance of the algorithm. Results are shown in \cref{fig:erdos}, with 20 random graphs considered per $n$. Taking $n=5$ as an example, the optimal vertex cover is found for all but 2 random instances tested. The solution quality decreases reasonably slowly, and for all trialled graphs the produced solution used at most 1.6 times the number of vertices as the optimal solution.

We also compare the performance of the quantum algorithm on minimum vertex cover to the best classical constant-factor approximation algorithm. The \textit{approximation factor} associated with an approximation algorithm is a proven guarantee on the ratio between the returned result and the optimal result. If this approximation factor is constant with respect to the problem size, the algorithm is referred to as a constant-factor approximation algorithm. This classical approximation algorithm for minimum vertex cover guarantees that the approximate cover will use at most twice as many nodes as the optimal cover for any input graph. It has time complexity $O(E)$, and is attributed to Fanica Gavril~\cite{Hartmanis:1982jy}, who discovered the algorithm in 1974.  After almost fifty years, no significant progress has been made in improving this 2-factor approximation. The current best-known approximation algorithm~\cite{Karakostas:2009ug} has an approximation factor bounded by $O(2 - \frac{1}{\sqrt{\log n}})$, converging to the same 2-factor approximation as the graph size increases. This classical algorithm has random result ratios in the range $[1.0, 2.0]$ on the same input graphs, independent of the size of the graph.

\begin{figure}[H]
\centering
\includegraphics[width=0.9\linewidth,valign=m]{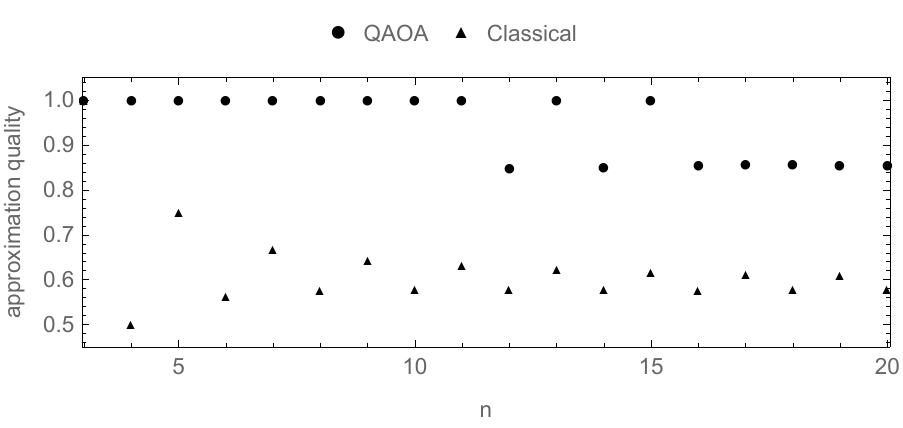}
\caption{Performance of the $p=2$ quantum algorithm on cycle graphs, with comparison to the classical $2$-approximation algorithm.}
\label{fig:results-circle}
\end{figure}

A cycle graph is a circle of $n \geq 3$ vertices connected by $n$ edges. The minimum vertex cover for a cycle graph must use $\ceil{n/2}$ vertices so that every edge is covered. From numerical tests, as $n$ becomes large the mean solution quality of the classical algorithm appears to converge to approximately $0.58$. As per \cref{fig:results-circle}, the $p=2$ algorithm outperforms this classical algorithm, with the solution quality above $80\%$ for all cycle graphs trialled.

A star graph of size $n$ has every vertex connected to a specific central vertex. For any star graph, the minimum vertex cover is the central vertex in the star graph, since all edges touch this vertex by definition. Star graphs are a useful group of graphs to study in the context of minimum vertex cover, because they are a pathological case for the classical approximation algorithm. A solution with two vertices is always chosen by this algorithm. Thus it always uses twice as many vertices as the minimum vertex cover on star graphs with $n \geq 2$. In contrast, QAOA performs exceedingly well on star graphs. With just $p=2$, the optimal solution is found for all trialled graphs having vertex count $n = 2, \ldots, 20$. In the context of the alternating operator perspective, a quantum walk of length $n-1$ to reach the 1-vertex state from the $n$-vertex initial state followed by amplification of this state via the honing operator is performed. This amplification of the optimal state is sufficiently large such that the single-vertex cover is discovered.

\begin{figure}[H]
\centering
\begin{subfigure}[b]{0.3\linewidth}
\includegraphics[width=\linewidth,valign=t]{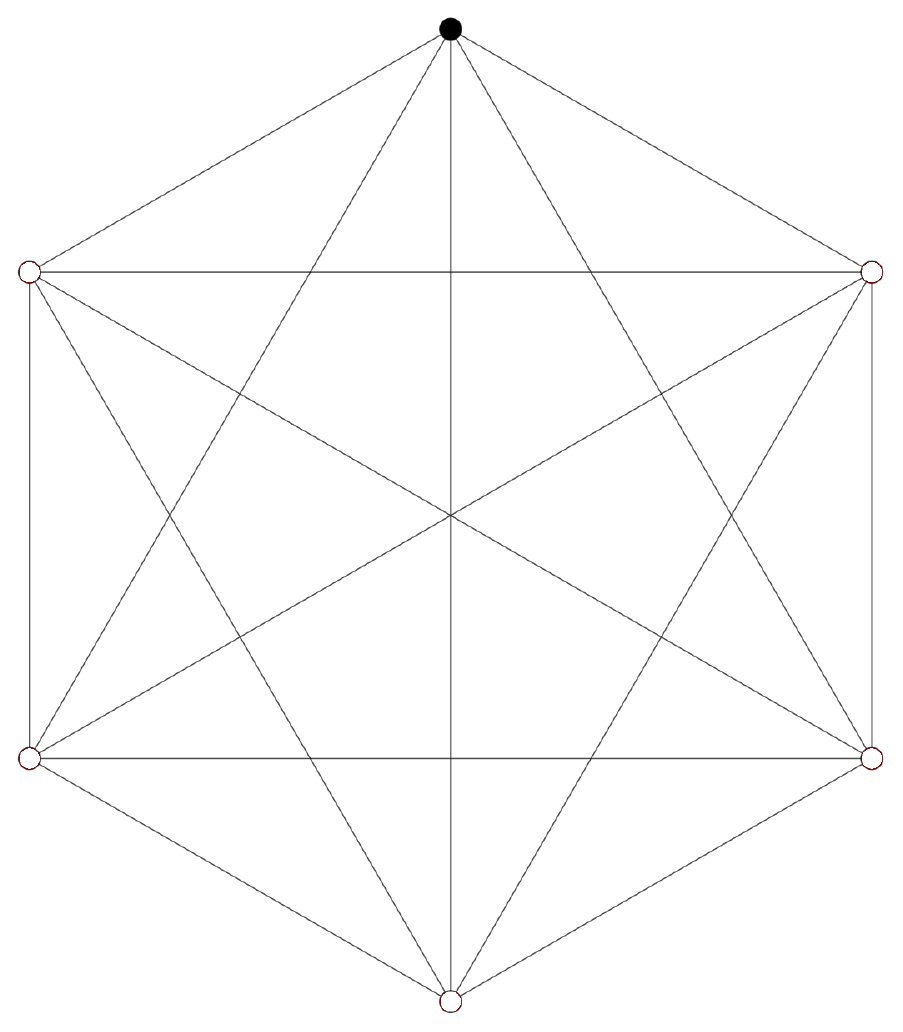}
\caption{$J(6, 1)$}
\label{fig:j61}
\end{subfigure}
\hfill
\begin{subfigure}[b]{0.3\linewidth}
\includegraphics[width=\linewidth,valign=t]{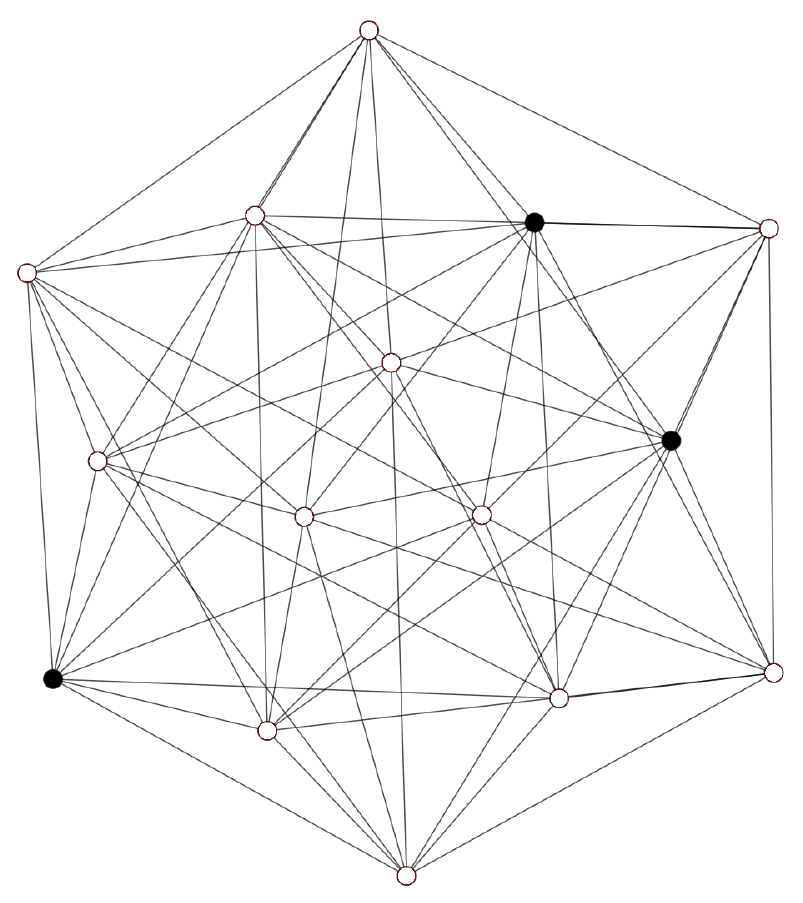}
\caption{$J(6, 2)$}
\label{fig:j62}
\end{subfigure}
\hfill
\begin{subfigure}[b]{0.3\linewidth}
\includegraphics[width=\linewidth,valign=t]{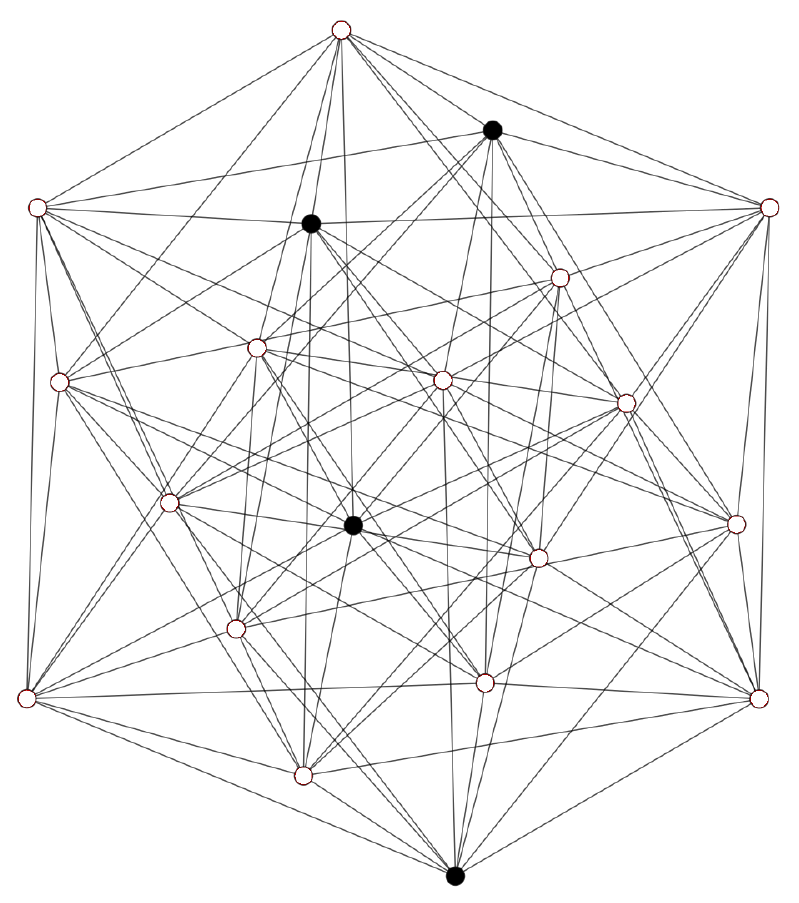}
\caption{$J(6, 3)$}
\label{fig:j63}
\end{subfigure}
\caption{Minimum vertex covers are shown for each $J(6, k=1,2,3)$ graph, indicated by white vertices. QAOA returns the optimal cover for each of these graphs, while the classical algorithm produces average solution qualities of $0.83$, $0.86$ and $0.85$ respectively.}
\label{fig:johnson}
\end{figure}

The Johnson graphs $J(n, k)$ have a number of properties which make them a useful family of graphs to study. The Johnson graph $J(n, k)$ has vertices labelled by each of the subsets of $\{1, 2, \ldots, n\}$ having size $k$. An edge connects two vertices if their intersection has size $k - 1$. The intrinsic difficulty they present in the graph isomorphism problem is also an open area of study~\cite{Babai2015}. Since graph isomorphism can be encoded into the Ising model~\cite{Lucas:2013eb} and has a bounded measure, it is valuable to look into the performance of the QAOA on Johnson graphs. Note also that $J(n, k)$ is isomorphic to $J(n, n-k)$, and that $J(n, 1)$ is a complete graph (every vertex is connected to every other vertex). The quality of approximations for $J(6, k)$ was evaluated, with $k = 1, 2, 3$. See \cref{fig:johnson} for the minimum vertex covers and performance of the classical versus the QAOA algorithm. QAOA performs optimally on these instances with $p=2$. Again, the quantum algorithm outperformed the classical algorithm, which produced sub-optimal solutions for each Johnson graph.

\section{Conclusion}

In this paper we have presented an algorithm for finding approximate solutions to NP optimisation problems with polynomially bounded measure (NPO PB) using the Quantum Approximate Optimisation Algorithm (QAOA). We have shown that the constraints involved with NP optimisation problems can be incorporated into the QAOA state evolution. This is done by interpreting the state evolution as a series of quantum walks, and then restricting the quantum walks to the region of feasible solutions. The QAOA also requires a method for efficiently finding the value of a certain expectation value. We have demonstrated that the recent concept of a hybrid quantum-classical variational algorithm suits for this purpose, and is efficient for NP optimisation problems that have polynomially bounded measure. Finally, the algorithm is applied to the NPO PB problem of minimum vertex cover. Classical simulations of the algorithm for graphs with up to $20$ vertices give promising results, using a fixed and low number of optimisation parameters.

There is significant potential for future work on various aspects of this QAOA-based algorithm. The algorithm supports any efficiently-preparable feasible solution as an initial state. Further work could investigate the impact of the choice of initial state, or even a superposition over multiple starting states -- in particular whether the `worst-case' solution (corresponding to the ground state of $\hat{C}$) is the best option for the initial state.

The graph used for quantum walks was modified from the transverse field (hypercube) operator, which is the conventional choice for an adiabatic evolution. However, other graphs could suit for this purpose. In terms of the continuous time quantum walk perspective, a quantum walk over any graph which connects the feasible states is a valid choice. Future work could investigate the impact of different choices for $\hat{B}$, or use a variety of different $\hat{B}$ in the quantum state evolution. 

Finally, applications to many other NP optimisation problems could be explored. Particular problems may produce symmetries in the quantum state evolution, leading to an expectation value which can be evaluated efficiently. Achieving this would remove the requirement for the variational sampling technique, and could potentially lead to a guarantee on the approximation factor. This has been done in the original QAOA paper for $p=1$ on the `maximum-cut' problem, but not for the modified version incorporating constraints. Analysing other problems in the NPO PB class and their corresponding state evolutions is a pathway for further research.

\bibliography{refs}

\end{document}